\documentclass[conference]{IEEEtran}
\IEEEoverridecommandlockouts
\usepackage{cite}
\usepackage{amsmath,amssymb,amsfonts}
\usepackage{algorithmic}
\usepackage{graphicx}
\usepackage{textcomp}
\usepackage{xcolor}
\usepackage{makecell}
\usepackage{multirow}
\usepackage{booktabs}
\usepackage{enumitem}
\def\BibTeX{{\rm B\kern-.05em{\sc i\kern-.025em b}\kern-.08em
    T\kern-.1667em\lower.7ex\hbox{E}\kern-.125emX}}
\begin{document}
\title{Reinforcement Learning-Based Pricing and Operation for Hybrid Electric and Hydrogen Vehicle Stations\\

    \thanks{L. Welburn is with the Mechanical Engineering Department, North Carolina A\&T State University, Greensboro, USA }
	\thanks{Y. Chen is with the Computational Sciences and Engineering Division, Oak Ridge National Laboratory, Oak Ridge, TN, USA (email: cheny4@ornl.gov)}
	\thanks{X. Wang is with the Department of Industrial and Systems Engineering, University of Tennessee at Knoxville, Knoxville, TN, USA}
	\thanks{A. Saroj is with the Buildings and Transportation Science Division, Oak Ridge, TN, USA }
    \thanks{S. Regev is with the Computational Sciences and Engineering Division, Oak Ridge National Laboratory, Oak Ridge, TN, USA }
	\thanks{This manuscript has been authored by UT-Battelle, LLC, under contract DE-AC05-00OR22725 with the US Department of Energy (DOE). The US government retains and the publisher, by accepting the article for publication, acknowledges that the US government retains a nonexclusive, paid-up, irrevocable, worldwide license to publish or reproduce the published form of this manuscript, or allow others to do so, for US government purposes. DOE will provide public access to these results of federally sponsored research in accordance with the DOE Public Access Plan (http://energy.gov/downloads/doe-public-access-plan)
	}
}

\author{Lowell Welburn, Yang Chen, Xudong Wang, Abhilasha Saroj, and Shaked Regev \vspace{-3mm}} 

\maketitle

\begin{abstract}
Electric vehicles (EVs) and hydrogen fuel cell vehicles (HFCVs) are promising technologies in the efforts to reduce carbon emissions. Challenges such as limited infrastructure, consumer dissatisfaction, and electrical grid strain inhibit progress in the adoption of these technologies. A zero emission vehicle (ZEV) station, capable of handling EVs and HFCVs, addresses these challenges by focusing on maximizing station profit and consumer utility, while minimizing grid impacts. The purpose of this research is to develop a realistic simulation of a ZEV station that is capable of handling real-time pricing decisions enabled by reinforcement learning (RL) and to compare the performance of the RL agent to static pricing methods. After training, the RL agent gains 10.83\% greater station profit than the baseline manual parameter search. This work serves as a useful tool for ZEV station owners when handling complex pricing and purchasing decisions.
\end{abstract}

\begin{IEEEkeywords}
Dynamic Pricing, Reinforcement Learning, Zero Emission Vehicle, Multi-Agent Simulation
\end{IEEEkeywords}

\section{Introduction}
Global climate change has been primarily attributed to an increase in carbon emissions over the last century \cite{dTing2021}. In 2022, light-, medium-, and heavy-duty vehicles within the transportation industry contributed 22.4\% of the total greenhouse gas (GHG) emissions in the US \cite{EPA2022}. Many efforts have been made to reduce carbon emissions in recent years, such as the adoption of zero emission vehicles (ZEVs) which include electric vehicles (EVs) and hydrogen fuel cell vehicles (HFCVs). Increasing popularity of ZEVs necessitates an increased infrastructure capacity to support these technologies. For this effort to be successful, the additional infrastructure must: 1) be financially viable, 2) maintain consumer trust and satisfaction, and 3) minimize negative impacts to power grid.

Currently, EV users most commonly choose between relatively slow AC charging from home and fast DC charging stations abroad. Residential time-of-use (TOU) programs commonly determine the price of electricity charged by utility providers depending on the time of day to incentivize customers to consume less electricity during high impact times, reducing strain on the energy grid. HFCV drivers use publicly or privately owned hydrogen fueling stations to refuel their vehicles, where the procedure and time duration is similar to refueling a gasoline vehicle \cite{eCelebi}.

The possibility of ZEV stations that are capable of handling the recharging and refueling for multiple vehicle fuel types has been explored, but further research is needed to support this effort \cite{yChen2025}. ZEV stations benefit from the ability to store and convert hydrogen to electricity or vice versa in times of need. ZEV stations could be more widely adopted to support multiple vehicle fuel types to improve the convenience of the charging and refueling process \cite{zevstation2026}.

Despite the growing need for integrated ZEV stations, existing pricing strategies proposed for real-world deployment by station owners largely rely on static models, traditional game theory, or rule-based heuristics that fail to adapt dynamically to real-time fluctuations in unexpected user demand and grid constraints. This inability to autonomously learn and optimize complex, multi-agent interactions often results in sub-optimal station profitability, increased consumer wait times, and unnecessary strain on the local power grid. Therefore, there is a critical need for an adaptive, data-driven pricing mechanism capable of simultaneously maximizing station profit, maintaining consumer utility, and minimizing grid impacts within a hybrid ZEV environment. Many previous works have focused on methods to dynamically set prices for the EV charging case, however, few studies have been performed for dynamic pricing of hybrid EV/HFCV charging/refueling stations where the complex relationships between the two systems are considered.

This paper presents the model of a multi-modal ZEV station along with a deep reinforcement learning (DRL) agent effective at making real-time pricing decisions for the ZEV station. The performance of the dynamic DRL agent is compared with manual, single-episode, and TOU static pricing methods. The primary contributions of this work are:
\begin{enumerate}[topsep=0pt,itemsep=0pt,parsep=0pt,partopsep=0pt]
    \item Development of a hybrid EV/HFCV multi-agent ZEV station simulation framework.
    \item Formulation of a PPO-based real-time pricing and energy management agent.
    \item Comparison between static, TOU, and fully dynamic RL pricing policies.
    \item Analysis of customer utility behaviors including balking and reneging under dynamic pricing.
\end{enumerate}

\section{Related Work}

\subsection{Traditional Pricing Strategies}
Strategies that do not rely on deep learning have been used in the past and currently for dynamic pricing. Game-theoretical models have been explored where local prosumer behavior and profit distribution were compared when working in cooperative and competitive scenarios \cite{yChen2020}. A multi-agent simulation (MAS) was presented where the charging station (CS) agent dynamically sets the price using a rule-based strategy built on the probabilistic demand distribution of arriving vehicles \cite{xWang2025}.

Multiple works have implemented Stackelberg game theoretic pricing. One of these uses a two-stage framework combining LSTM demand forecasting with a bilevel Stackelberg game optimization to determine both the location and pricing of charging stations, where the leader selects locations and followers set prices to maximize payoffs \cite{mAdil2024}. Another introduces a conic relaxation method for solving the non-convex Stackelberg game between profit-maximizing charging station operators and cost-minimizing EV users whose routing choices are influenced by pricing decisions \cite{hZeng2024}. Strategic interaction between HFCVs and hydrogen refueling stations has also been modeled using a Stackelberg game approach where the proposed pricing and refueling strategy simultaneously improves station profit and reduces consumer refueling costs \cite{hWang2024}. Several efforts have also focused on optimization-based pricing techniques \cite{dMatkovic2025,dLiu2023,dLu2023}. A bilevel model has been applied where a charging network provider (leader) sets time-varying prices across stations to maximize profit and reduce grid impact, while customers (followers) independently choose charging times subject to those prices \cite{mAnjos2025}. Many of the previously mentioned strategies lack the ability to autonomously adapt to a changing world environment and do not scale well to large multi-agent simulations which is why deep learning methods have been studied recently.

\subsection{Dynamic Pricing Using Deep Learning}
With the rise of artificial intelligence (AI) and machine learning (ML), deep learning has been employed is various research efforts using unique approaches to solve the problems with dynamic pricing and negative grid impacts \cite{qChen2023}. One such work integrated real-time data with an XGBoost-based algorithm to develop a day-ahead dynamic pricing model to help reduce grid impacts \cite{bPalaniyappan2024}. Another work presented a novel Multi-Agent Deep Neural Network (MADNN) that considers energy balance arrival time, departure time, and range anxiety to determine dynamic pricing of EV fast chargers \cite{bAljafari2023}.

A variety of DRL-specific methods have also been implemented, helping to solve different aspects of the dynamic charging problem. For example, DRL has been successfully used to help local power grids solve charging scheduling problems to reduce uneven load distribution and provide customers with lower prices \cite{dLiu2023,tQi2023}. Narayan et al. presented a method to maximize an aggregator's revenue based on a Deep Q-Network (DQN) RL agent that has pricing control over multiple EV charging stations throughout the environment \cite{aNarayan2022}. Another study showed the use of DRL to vehicle routing and energy scheduling of emergency EVs during events where the power supply is affected \cite{mAlqahtani2022}. A novel DRL approach that maximizes the quality of service was also developed and validated \cite{zZhao2022}. The soft actor-critic (SAC) reinforcement learning algorithm has been implemented to determine the optimal spatial placement and dynamic pricing of EV charging stations \cite{yLi2023}. Cui et al. presented a Deep Deterministic Policy Gradient (DDPG) based agent that aimed to maximize charging station profit through dynamic pricing in a competitive scenario built on a traffic flow prediction model \cite{lCui2023}.

A limited number of studies have focused on a multi-functional charging station for electric and hydrogen vehicles. One such work focuses on the relationship between hydrogen production facilities and the multi-purpose stations that relies on an upper level Stackelberg game approach for hydrogen trading and a lower level  asymmetric Nash bargaining game for electricity trading between charging stations \cite{cWei2026}. Another study compares multiple DRL algorithms that optimized the retail price of hydrogen for a hydrogen refueling station that also interacts with the electricity grid \cite{bLi2023}.

\section{Methodology}
\begin{figure}[t]
\centerline{\includegraphics[width=0.85\columnwidth]{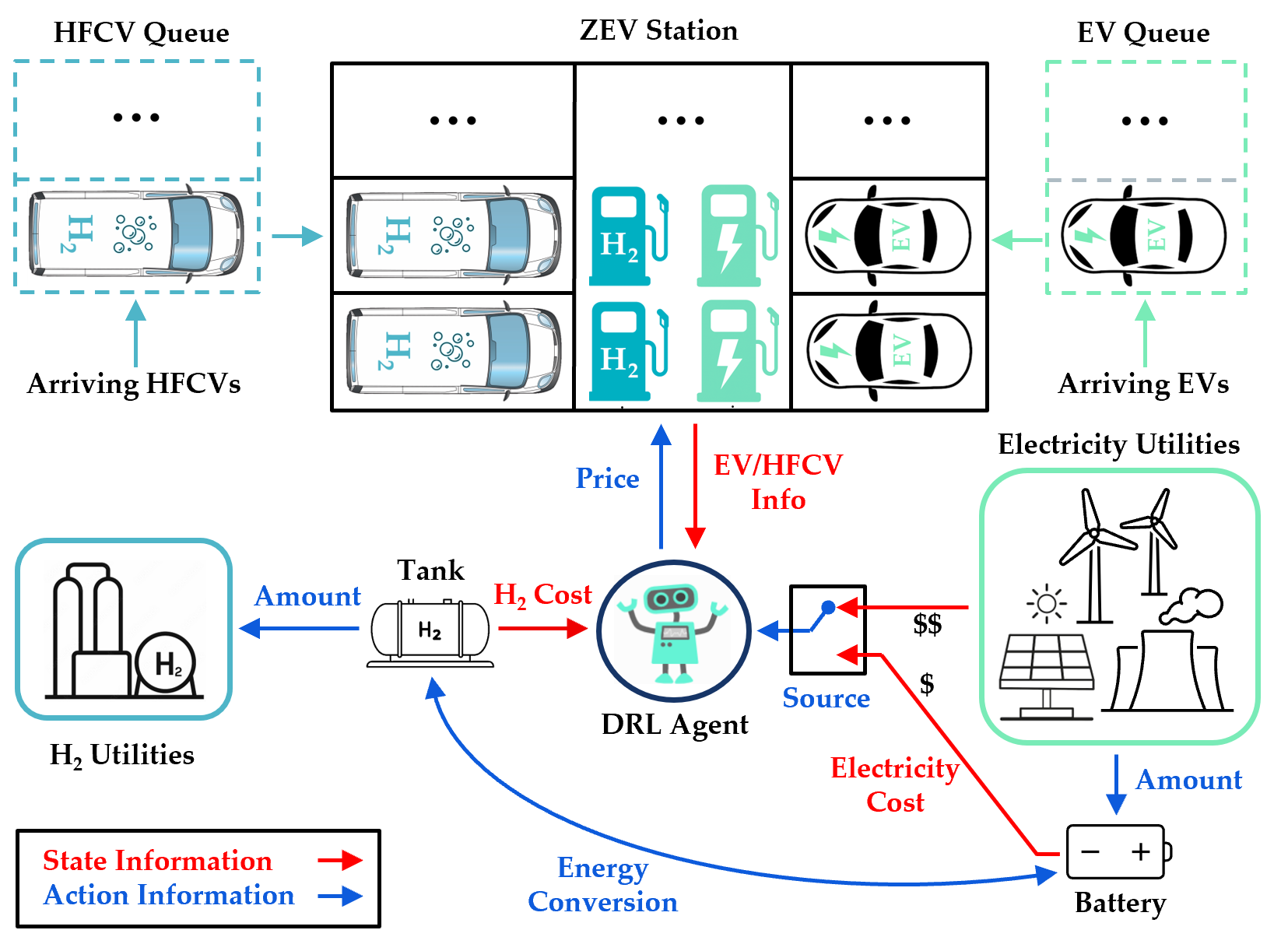}}\
\centerline{\includegraphics[width=0.9\columnwidth]{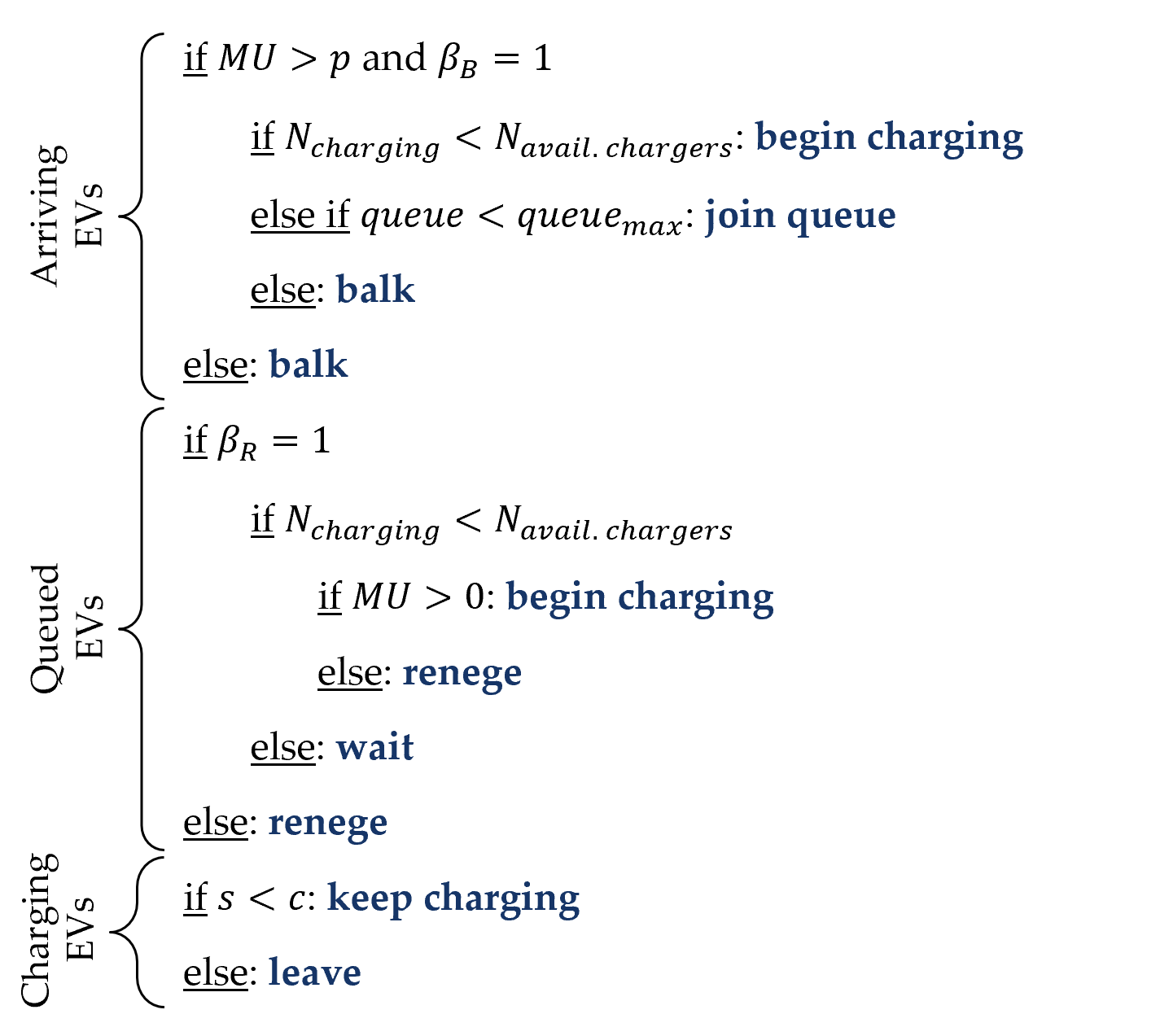}}
\caption{ZEV station setup outlining DRL agent action space and vehicle state logic.}
\label{fig1}
\end{figure}

\begin{figure}[t]
\centerline{\includegraphics[width=\columnwidth]{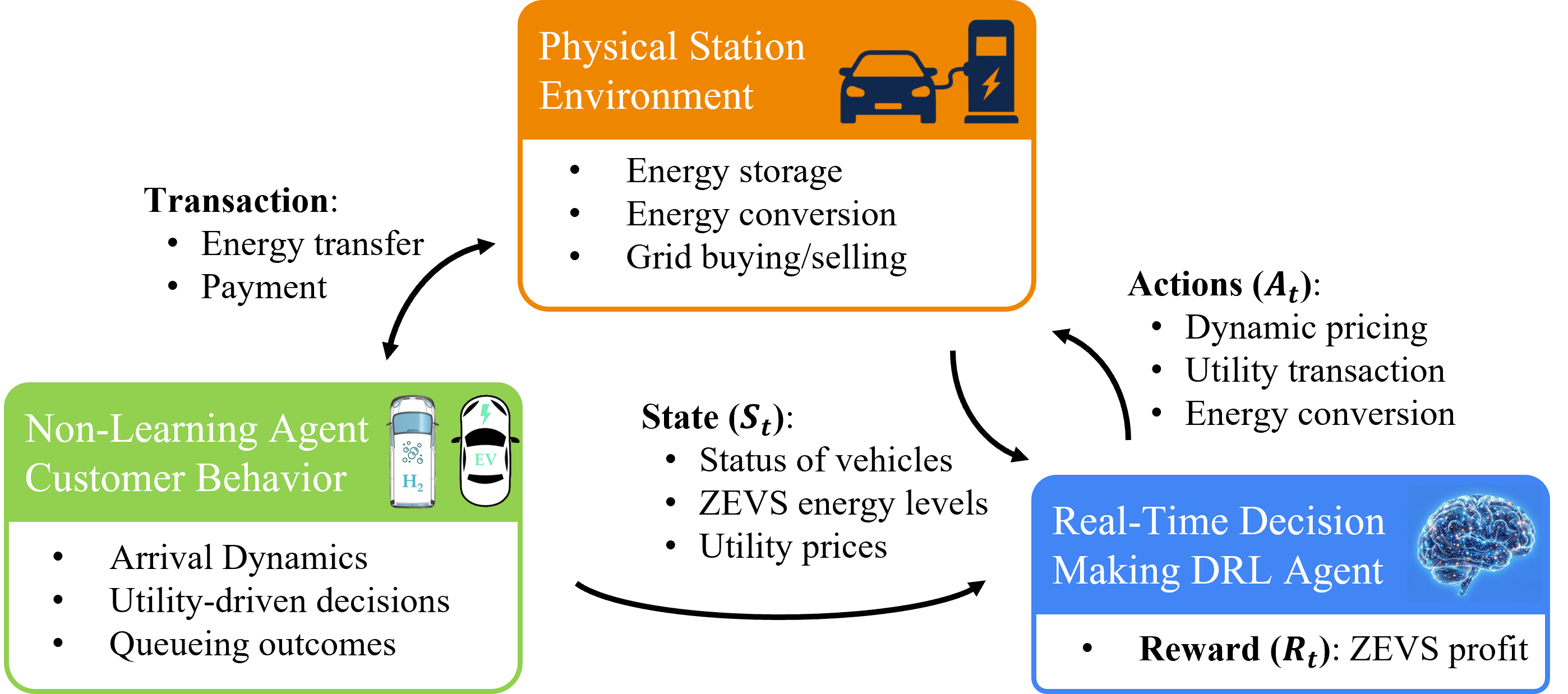}}
\caption{Hybrid ZEV station concept framework.}
\label{fig2}
\end{figure}

The present work builds off of prior work research that integrated many dynamic factors such as battery storage, hydrogen production, hydrogen storage, fuel cells, and dynamic interactions with the electricity grid \cite{jGbadegoye2025}. The simulation consists of multiple rule-based non-learning EV and HFCV agents and a single DRL agent dynamically making pricing, purchasing, and energy conversion decisions, as shown in Fig.~\ref{fig1}. The ZEV station has a battery storage of 2000 kWh and a hydrogen tank storage of 500 kg. For this simulation, energy can be transferred from electricity to hydrogen and vice versa via electrolyzer and fuel cell capabilities, with both having an assumed efficiency of 50\% \cite{yChen2025}. Each episode runs for 24 hours, with a timestep size of 5 minutes. For this simulation the TOU rates are defined as: 1) discount from 1am-6am, 2) off-peak from 6am-6pm and 9pm-1am, and 3) peak from 6pm-9pm. Fig.~\ref{fig2} shows the concept framework interactions between the DRL agent, physical station environment, and non-learning customer agents.

\subsection{Non-Learning Agents}
Random variance in daily incoming customer arrival rates is modeled by EV and HFCV agents arriving at the simulated station stochastically according to Poisson distributions, which varies depending on the time of day. Equation \eqref{eq1} is used to calculate the incremental utility gain of an EV or HFCV user, where \eqref{eq2} accounts for the state of charge \cite{sLai2024}. The marginal utility for consuming one more unit of recharging/refueling goods is then calculated using \eqref{eq3}.
\vspace{-2pt}
\begin{equation}
U^t=U^{t-1}+\alpha \log(v\Delta t/E+1)\theta-v\Delta tp-\sigma t\label{eq1}
\end{equation}
\begin{equation}
\theta=e^{-(-\ln(Pr))^{1-x}}\label{eq2}
\end{equation}
\begin{equation}
M^t=\frac{U^t-U^{t-1}}{s^t-s^{t-1}}\label{eq3}
\end{equation}

For the EV case, $U^t$ is the utility at time step $t$, $\alpha$ is the coefficient representing the perceived benefit from charging, $v$ is the charging rate, $\Delta t$ is the time increment, $E$ is the battery capacity, $p$ is the energy price, and $\sigma$ is the waiting cost coefficient. Generally, the consumer gains positive utility from charging and negative utility from price and waiting time at the station. $\theta$ is the state-of-charge weighting factor defined in \eqref{eq2}, where $Pr$ is the probability of receiving utility from energy gained, and $x$ is the state of charge. As the state-of-charge decreases, the consumer is more likely to purchase the fuel. $M^t$ represents the marginal utility of additional energy consumption where $s$ is the energy level.

When the vehicles arrive at the ZEV station, they can also view the queue length ($N_q$) and number of vehicles at the station ($N_z$). Balking occurs when a vehicle arrives at the station but refuses to charge or join the queue, while reneging is defined as a vehicle choosing to leave the queue before refueling. The EV and HFCV probabilities for balking and reneging are given by \eqref{eq4} and \eqref{eq5}, where $\gamma$ is the probability decrease rate parameter and $\delta$ is the non-negative adjusting factor \cite{sLai2023}. These parameters determine the magnitude of the effect that queue length and number of vehicles at the station have on the probabilities for balking and reneging, respectively.

\vspace{-2pt}
\begin{equation}
\mathbb{P}_B=1-e^{-N_q\gamma}\label{eq4}
\end{equation}
\begin{equation}
\mathbb{P}_R=\delta\ln\frac{N_q+1}{N_z}\label{eq5}
\end{equation}

Vehicles progress from arrival to recharging/refueling if the user marginal utility is greater than the current price of electricity/hydrogen and the probability of balking condition is met. If the user marginal utility condition is met and all of the charging/fueling spots are occupied, the vehicle will join the queue if space is available.
\vspace{-3pt}
\begin{table}[b]
\caption{PPO Hyperparameter Settings}
\label{tab1}
\centering
\begin{tabular}{lc}
\toprule
\textbf{Hyperparameter} & \textbf{Value} \\
\midrule
Learning rate ($\alpha$) & $0.0003$ \\
Horizon ($T$) & $2048$ \\
Minibatch size ($M$) & $64$ \\
Discount factor ($\gamma$) & $0.99$ \\
GAE parameter ($\lambda$) & $0.95$ \\
Clip range ($\epsilon$) & $0.2$ \\
Value function coefficient ($c_1$) & $0.5$ \\
Entropy coefficient ($c_2$) & $0.0$ \\
Number of epochs & $10$ \\
Total training timesteps & $1 \times 10^6$ \\
\bottomrule
\end{tabular}
\end{table}
\vspace{-2pt}
\subsection{Learning Agent}
The Proximal Policy Optimization (PPO) algorithm framework is chosen to implement the DRL agent because of its relatively stable training and insensitivity to hyperparameter adjustments. As part of the state space the agent can view the normalized values of the number of vehicles in the queue ($N_q$), number of refueling vehicles ($N_r$), served vehicle ratio ($V_s$), ZEV station battery/fuel level ($B_z$), consumer refueling demand ($d$), and market price of utilities ($p_u$). The action space includes the dynamic price ($p$), amount of utilities to purchase ($d_u$), amount of energy conversion ($C_z$), and amount of electricity to sell back to the market ($S_m$). The PPO algorithm was trained for $1e6$ timesteps and initialized with hyperparameter values found in Table~\ref{tab1}.

Equations \eqref{eq6}-\eqref{eq9} show the reward ($R$) that the agent receives throughout each episode, which tracks the net profit ($P$) from EVs and HFCVs per timestep, which includes net gains ($P_g$), net losses ($P_l$), number of served vehicles ($N_s$), number of arrived vehicles ($N_a$), average electricity demand ($d_{avg}$), current electricity demand ($d_{ev}$), and theoretical maximum electricity demand ($d_{max}$). Weights $\omega_1$, $\omega_2$, and $\omega_3$ are adjusted manually via trial and error depending on the ZEV station preferences for station profit versus social welfare. The profit reward ($R_p$) is maximized as gains are increased and losses are decreased. The served ratio reward ($R_s$) increases as prices remain below consumer utility thresholds, incentivizing customers to refuel. The demand reward ($R_d$) is optimized as the difference between the current timestep demand and the running average demand is minimized. The number of served vehicles increases when a vehicle has progressed successfully from arrival to completing the charging/refueling process.

\begin{equation}
R_p=P_{g}-P_{l}\label{eq6}
\end{equation}
\begin{equation}
R_s=N_{s}/N_{a}\label{eq7}
\end{equation}
\begin{equation}
R_d=-|d_{avg}-d_{ev}|/d_{max}\label{eq8}
\end{equation}
\begin{equation}
R=\omega_1 R_p+\omega_2 R_s+\omega_3 R_d\label{eq9}
\end{equation}

The agent is compared against a manual setup and is trained under four conditions: 1) Static, 2) TOU, and 3) Dynamic Case 1, and 4) Dynamic Case 2.

\begin{itemize}
\item \textbf{Manual} - the price is set based on manual parameter search over a range of fixed prices.
\item \textbf{RL Static} - the agent can only select one electricity and hydrogen price at the start of each episode.
\item \textbf{RL TOU} - the agent can select one electricity and hydrogen price at the start of each episode and with every TOU scheduling shift (5 actions per episode).
\item \textbf{RL Dynamic: Case 1} - the agent can select one electricity and hydrogen price at every timestep of the episode and focuses only on maximizing station profit by setting $\omega_1=1$, $\omega_2=0$, and $\omega_3=0$ in \eqref{eq9}.
\item \textbf{RL Dynamic: Case 2} - the agent can select one electricity and hydrogen price at every timestep of the episode and focuses on balancing station profit, customer utility, and grid impact by setting $\omega_1=1$, $\omega_2=0.01$, and $\omega_3=0.01$ in \eqref{eq9}.
\end{itemize}

\section{Results}
\begin{figure}[t]
\centerline{\includegraphics[width=\columnwidth]{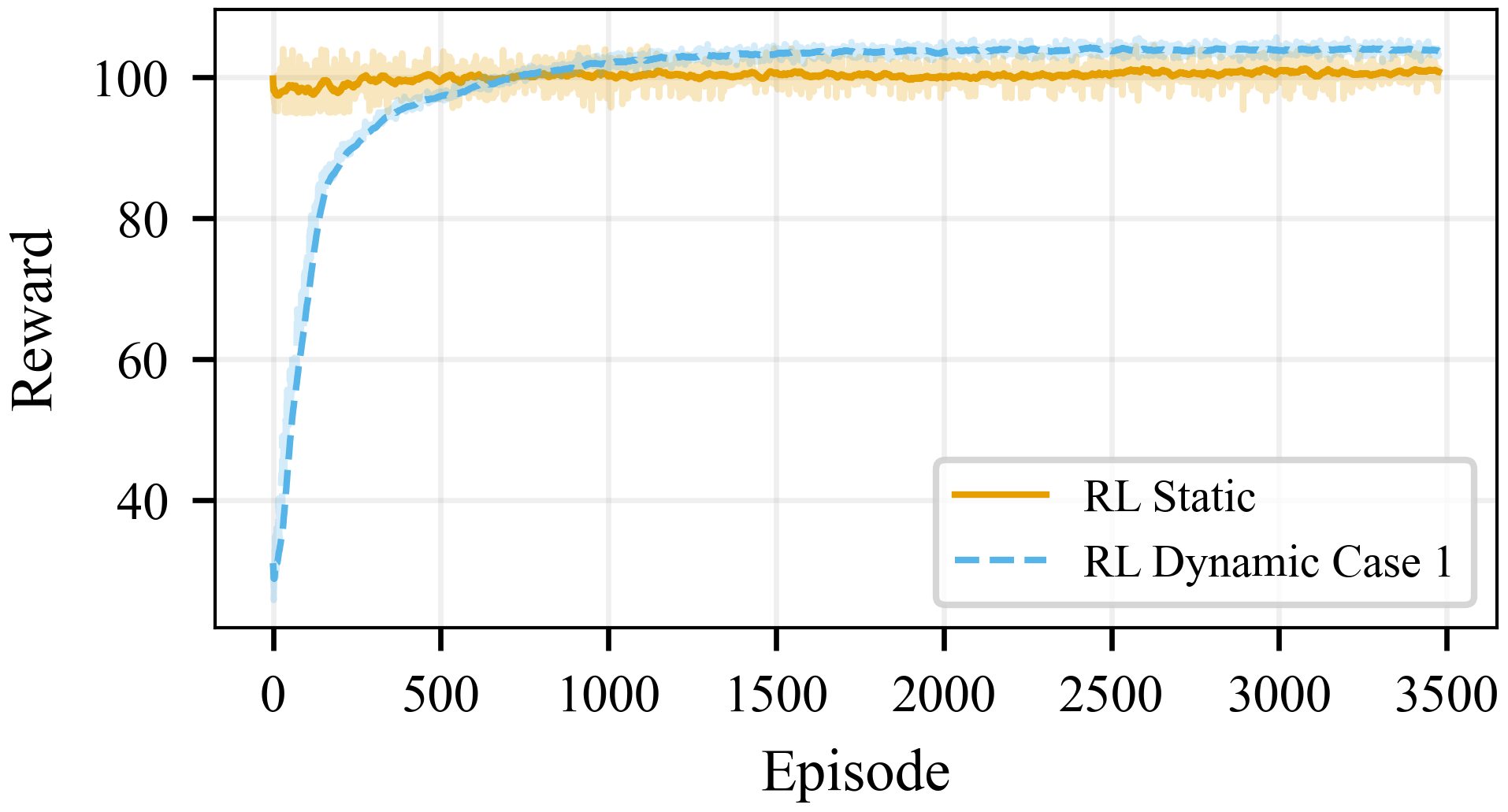}}
\caption{Mean training reward comparison.}
\label{fig3}
\end{figure}

Initially the static training produces higher reward due to price boundary action space limitations; however, the dynamic training quickly surpasses the performance of the former, as seen in Fig.~\ref{fig3}. One explanation for the steep learning of the dynamic agent is the frequency of action-observation pairs. At each timestep, the agent is able to choose a new action and receive a state observation. For the static case, the agent is limited to choosing one action at the beginning of the episode, and then for the remainder of the timesteps can only observe the state, delaying the time between actions and their consequences. This produces an ill-posed Markovian decision-making process, where credit assignment is difficult to ascertain.

\begin{figure}[!b]
\centerline{\includegraphics[width=\columnwidth]{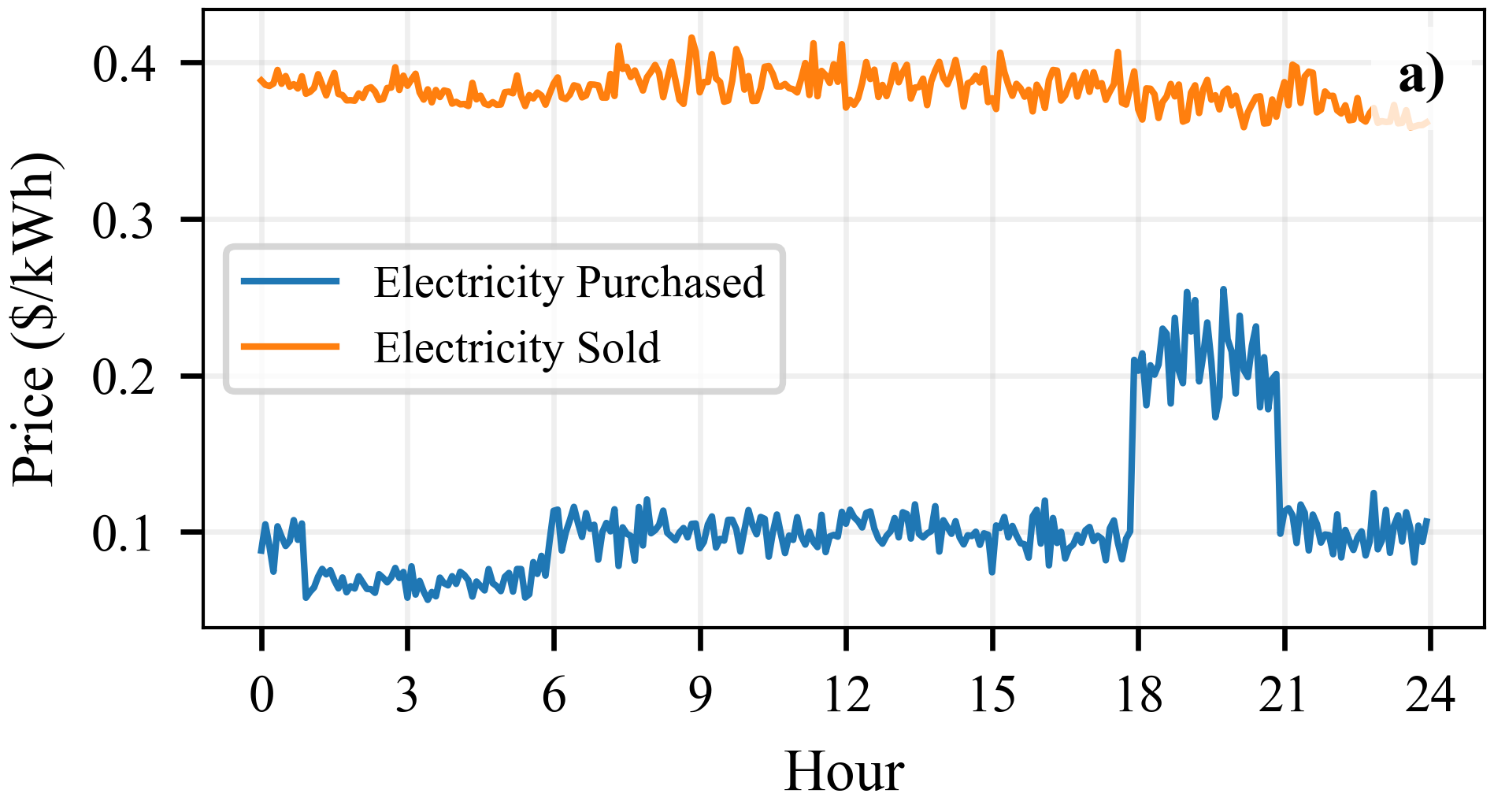}}
\centerline{\includegraphics[width=\columnwidth]{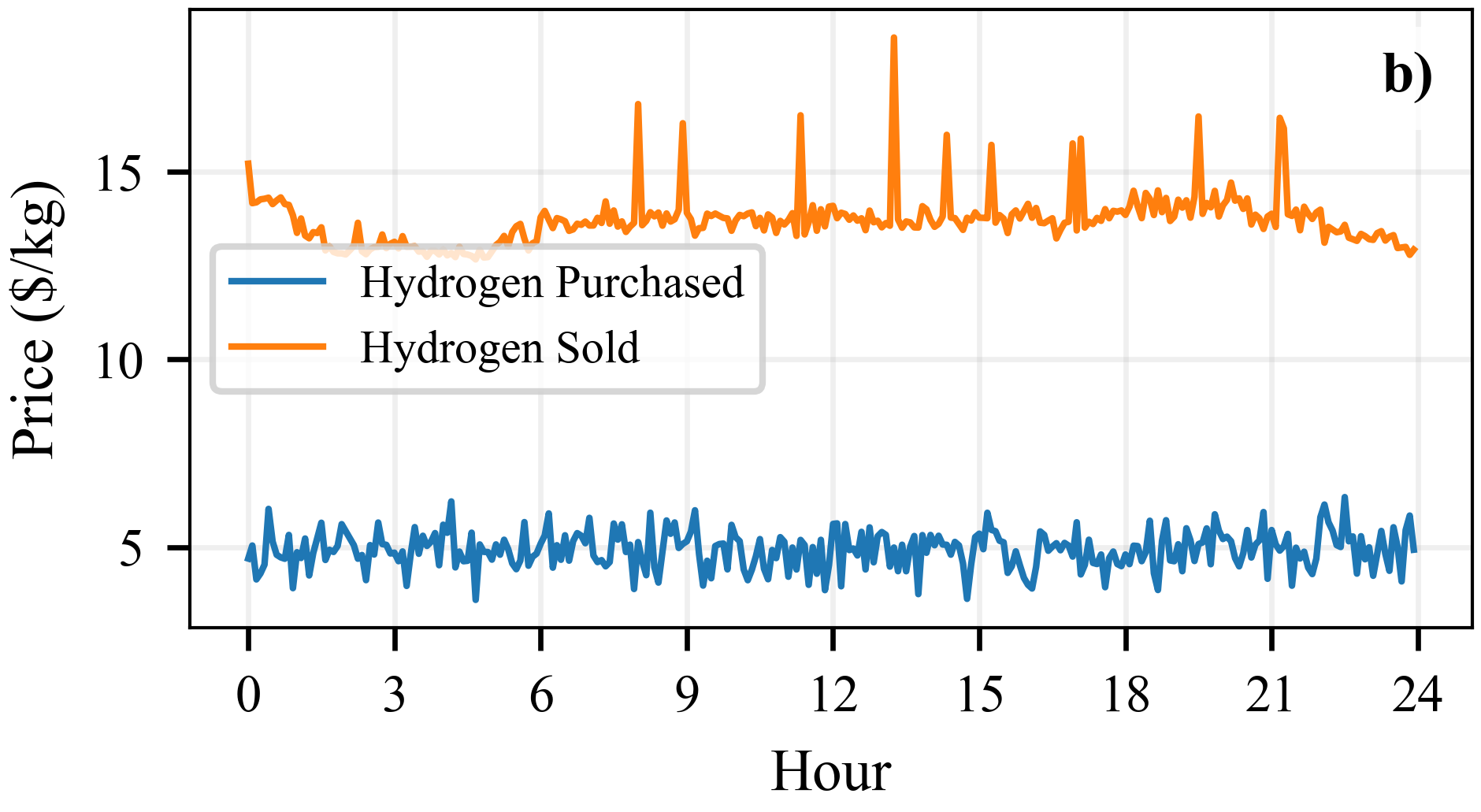}}
\caption{Dynamic market prices and ZEV selling prices.}
\label{fig4}
\end{figure}

\vspace{-2pt}
Fig.~\ref{fig4} shows the market and selling price of electricity and hydrogen throughout a validation episode in the RL Dynamic Case 1. The agent chooses to raise the selling price generally as the TOU rates progress from discount to off-peak (6am) and on-peak rates (6pm).

Table~\ref{tab2} provides a comparison of the results using the four methods discussed earlier. The net profit is computed as a mean including each of the validation episodes and the price is calculated using a volume-weighted averaging method. The fueled, balk, and renege percentages are averaged over the same validation episodes. Compared with the Manual method, the RL Static, RL TOU, and RL Dynamic produced a -27.13\%, -4.36\%, +10.83\%, and +8.27\% change in net profit respectively, and a +15.83\%, +7.27\%, +11.02\%, and +6.63\% change in served customer percentage respectively. This shows that the RL Dynamic method produces the highest profit, while also showing promise for increasing consumer utility. The RL Dynamic agent in both cases is able to perform higher partly because it has the ability to see a low market price in real-time and choose to buy from the grid during these low-price periods.

\begin{figure}[t]
\centerline{\includegraphics[width=\columnwidth]{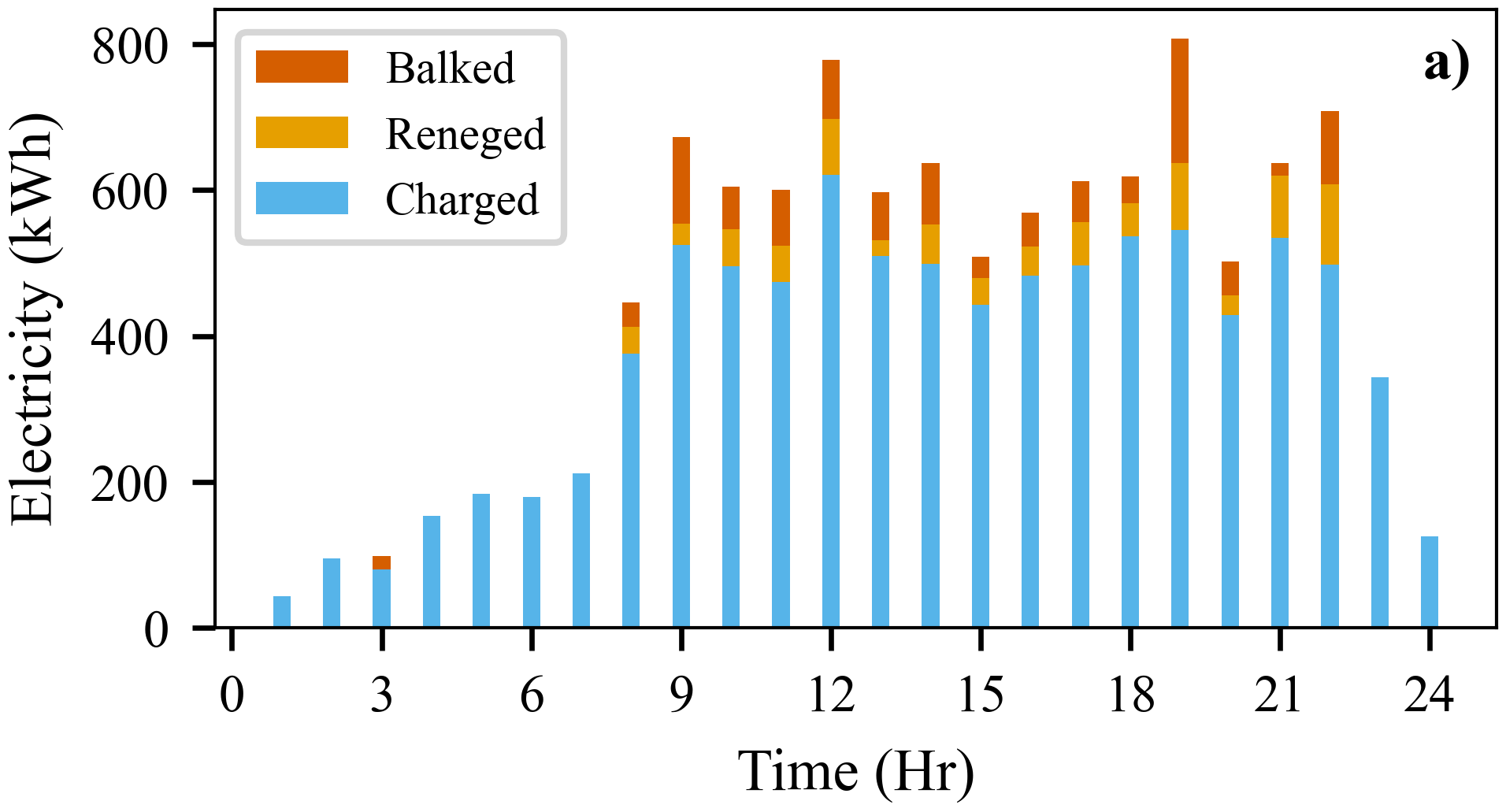}}
\centerline{\includegraphics[width=\columnwidth]{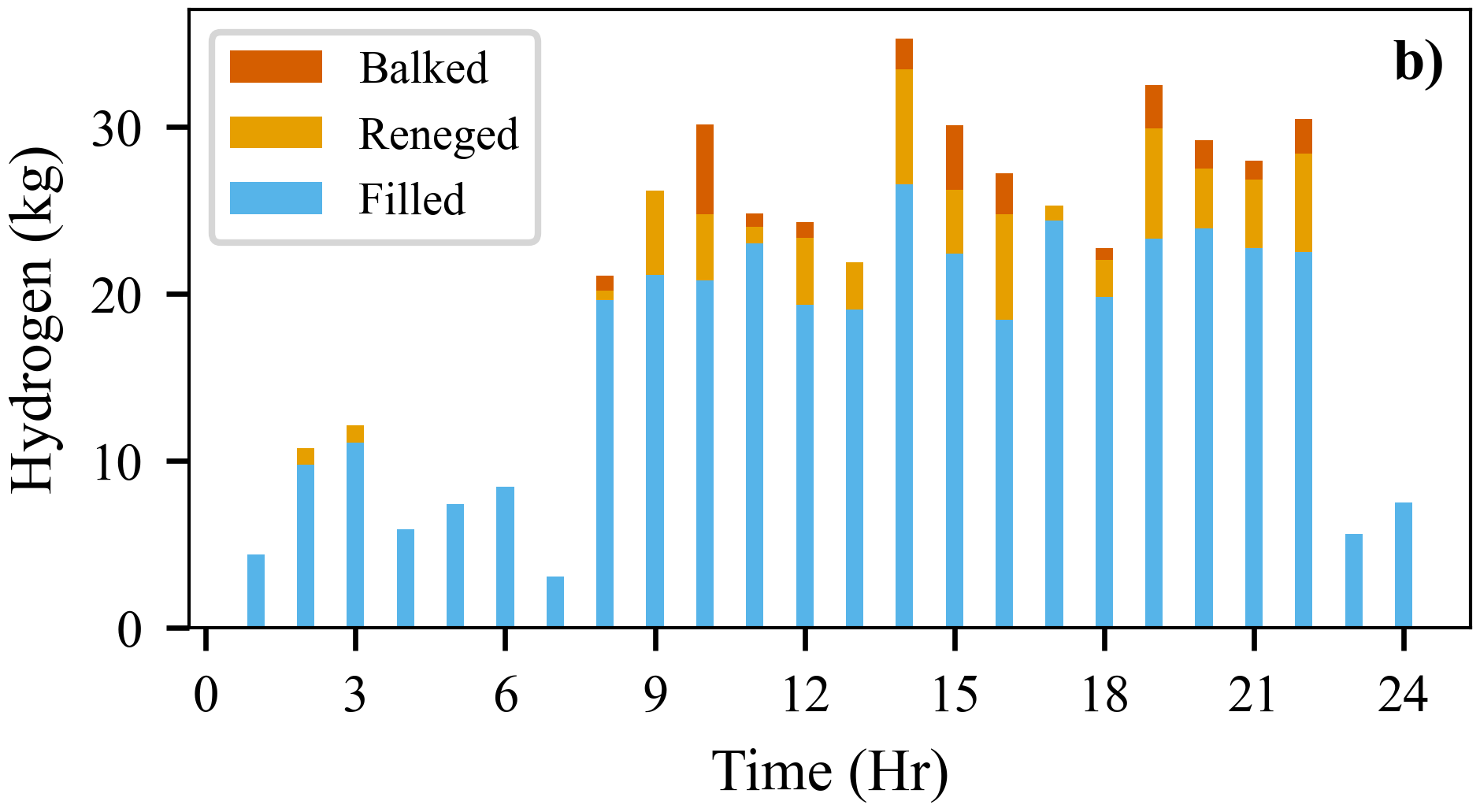}}
\caption{Amount of energy transaction lost due to balking and reneging and amount of energy a) charged for EVs and b) filled for HFCVs in the RL dynamic case.}
\label{fig5}
\end{figure}

The balking and reneging trends can be seen in Fig.~\ref{fig5}, where each bar represents the aggregated data from the previous hour. The highest rates of balking and reneging align with daytime hours where more vehicles are arriving at the station. From the results, the primary contributor to balking and reneging is the number of vehicles currently in the queue, which is also reflected theoretically in \eqref{eq4}-\eqref{eq5}.

\vspace{-2pt}

\begin{figure}[b]
\centerline{\includegraphics[width=\columnwidth]{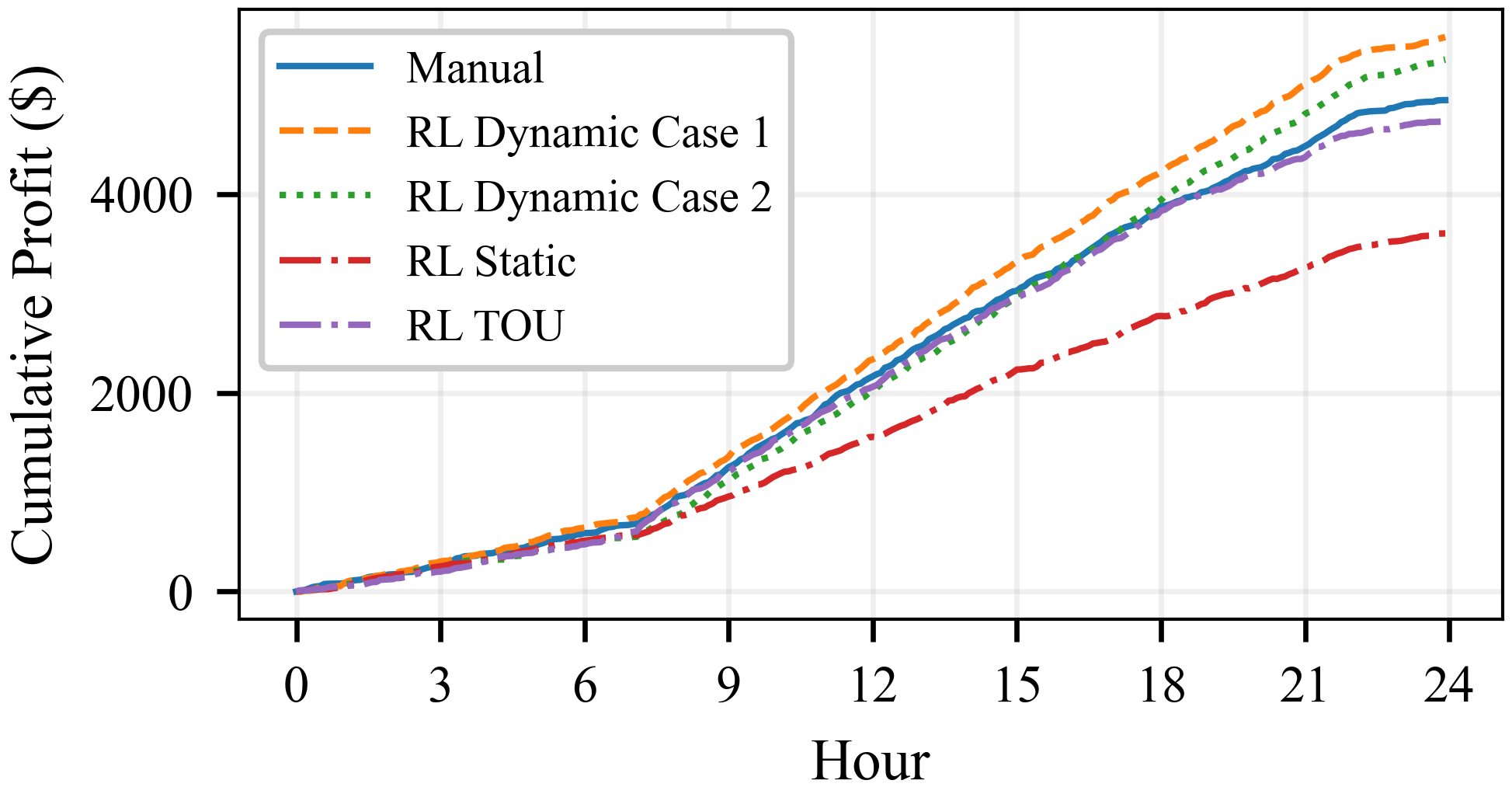}}
\caption{Averaged ZEV station net profit comparison over validation episodes.}
\label{fig6}
\end{figure}

Furthermore, the cumulative financial performance of the various pricing strategies is illustrated in Fig. \ref{fig6}, which displays the averaged ZEV station net profit over the validation episodes. The RL Dynamic agent consistently maintains a steeper profit accumulation trajectory compared to the Manual and RL Static methods. The RL TOU agent performs comparably to the RL Dynamic agent, suggesting that adapting to the utility company's TOU schedule captures a significant portion of the available profit. However, the RL Dynamic agent's ability to make micro-adjustments at every timestep allows it to exploit minor market inefficiencies, ultimately yielding the highest final profit.

As detailed in Table \ref{tab2}, the station generated significantly higher net profits from HFCV users compared to EV users across all methods. This disparity can be attributed to the differing market dynamics and consumer utility profiles of the two energy types, reflected in different coefficient weights between vehicle types used for perceived benefit of refueling found in \eqref{eq1}. HFCV users generally possess a higher tolerance for elevated prices due to the scarcity of alternative hydrogen refueling infrastructure, allowing the agent to sustain a higher volume-weighted average selling price of 14.42 \$/kg. Conversely, the high availability of home EV charging places a lower ceiling on the maximum price the agent can enforce before EV consumers balk or renege.

\begin{table}[t]
\centering
\renewcommand{\arraystretch}{1.2}
\setlength{\tabcolsep}{4pt} 
\caption{Performance comparison between chosen methods over validation episodes}
\begin{tabular}{ccccccc}
\toprule
\textbf{Type} & \textbf{Method} & \textbf{\makecell{Net \\ Profit \\ (\$)}} & \textbf{\makecell{Price \\ (\$/kWh) \\ or (\$/kg)}} & \textbf{\makecell{Fueled \\ (\%)}} & \textbf{\makecell{Balk \\ (\%)}} & \textbf{\makecell{Renege \\ (\%)}} \\
\midrule
\multirow{4}{*}{EV} & Manual & 2166.83 & 0.40 & 81.76 & 10.17 & 8.07\\
& RL Static & 2208.36 & 0.42 & 73.65 & 20.47 & 5.88\\
& RL TOU & 1284.77 & 0.28 & 81.53 & 8.68 & 9.79\\
& RL Dyn. Case 1 & 2131.99 & 0.39 & 82.75 & 9.68 & 7.57\\
& RL Dyn. Case 2 & 2240.18 & 0.40 & 79.32 & 12.09 & 8.59\\
\midrule
\multirow{4}{*}{HFCV} & Manual & 2778.61 & 15.00 & 66.90 & 30.83 & 2.26\\
& RL Static & 1395.30 & 10.00 & 98.54 & 0.40 & 1.06\\
& RL TOU & 3445.17 & 14.75 & 77.94 & 10.95 & 11.11\\
& RL Dyn. Case 1 & 3348.81 & 14.16 & 82.29 & 5.13 & 12.59\\
& RL Dyn. Case 2 & 3114.12 & 13.77 & 79.19 & 7.00 & 13.81\\
\bottomrule
\end{tabular}
\label{tab2}
\end{table}

Despite the availability of energy conversion technologies (electrolyzer and fuel cell) within the ZEV station model, analysis of the logged agent actions during the validation episodes indicates that the RL Dynamic agent strictly abstained from utilizing the energy conversion action ($C_z$). This learned behavior is a direct consequence of the severe economic penalty imposed by the 50\% efficiency loss associated with these cross-conversion processes.

\section{Conclusion}
This paper presented a multi-agent simulation framework for a hybrid ZEV station capable of servicing both EVs and HFCVs. A DRL agent utilizing the PPO algorithm was developed to make dynamic pricing decisions and the performance of the continuous DRL agent was evaluated against manual, static, and TOU pricing strategies. The results demonstrate that the dynamic agent achieved a 10.83\% increase in net station profit compared to manual pricing, while simultaneously increasing the served customer percentage by +11.02\%. Analysis of consumer balking and reneging behaviors highlighted that queue length during peak daytime hours is the primary driver of lost transactions. Future work could benefit from validation against real-world charging station data. Ultimately, this DRL-based dynamic pricing framework provides an autonomous tool for ZEV station operators to maximize profitability, adapt to fluctuating market conditions, and support the broader integration of sustainable transportation infrastructure.

\section*{Acknowledgment}
The first author gratefully acknowledges support from the National Nuclear Security Administration (NNSA) Minority Serving Institution Partnership Program (MSIPP) during 2025--2026.

\bibliographystyle{IEEEtran}
\bibliography{references.bib}

@article{EPA2022,
    author = {EPA},
    year = {2025},
    title = {Fast Facts on Transportation Greenhouse Gas Emissions},
    journal = {https://www.epa.gov/greenvehicles/fast-facts-transportation-greenhouse-gas-emissions}
}

@ARTICLE{eCelebi,
  author={Celebi, Emre and Fuller, J. David},
  journal={IEEE Transactions on Power Systems}, 
  title={Time-of-Use Pricing in Electricity Markets Under Different Market Structures}, 
  year={2012},
  volume={27},
  number={3},
  pages={1170-1181},
  doi={10.1109/TPWRS.2011.2180935}}

@article{dTing2021,
    author = {Ting, David and Vasel-Be-Hagh, Ahmad},
    year = {2021},
    title = {Mitigating Climate Change},
    journal = {Proceedings of the Mitigating Climate Change 2021 Symposium and Industry Summit}
}

@Article{dLiu2023,
    AUTHOR = {Liu, Ding and Zeng, Peng and Cui, Shijie and Song, Chunhe},
    TITLE = {Deep Reinforcement Learning for Charging Scheduling of Electric Vehicles Considering Distribution Network Voltage Stability},
    JOURNAL = {Sensors},
    VOLUME = {23},
    YEAR = {2023},
    NUMBER = {3},
    ARTICLE-NUMBER = {1618},
    URL = {https://www.mdpi.com/1424-8220/23/3/1618},
    PubMedID = {36772657},
    ISSN = {1424-8220},
    DOI = {10.3390/s23031618}
}

@ARTICLE{tQi2023,
    author={Qi, Taoyi and Ye, Chengjin and Zhao, Yuming and Li, Lingyang and Ding, Yi},
    journal={Journal of Modern Power Systems and Clean Energy}, 
    title={Deep Reinforcement Learning Based Charging Scheduling for Household Electric Vehicles in Active Distribution Network}, 
    year={2023},
    volume={11},
    number={6},
    pages={1890-1901},
    doi={10.35833/MPCE.2022.000456}
}

@article{mAlqahtani2022,
    title = {Dynamic energy scheduling and routing of multiple electric vehicles using deep reinforcement learning},
    journal = {Energy},
    volume = {244},
    pages = {122626},
    year = {2022},
    issn = {0360-5442},
    doi = {https://doi.org/10.1016/j.energy.2021.122626},
    url = {https://www.sciencedirect.com/science/article/pii/S0360544221028759},
    author = {Mohammed Alqahtani and Mengqi Hu}
}

@article{yLi2023,
    title = {Dynamic pricing based electric vehicle charging station location strategy using reinforcement learning},
    journal = {Energy},
    volume = {281},
    pages = {128284},
    year = {2023},
    issn = {0360-5442},
    doi = {https://doi.org/10.1016/j.energy.2023.128284},
    url = {https://www.sciencedirect.com/science/article/pii/S036054422301678X},
    author = {Yanbin Li and Jiani Wang and Weiye Wang and Chang Liu and Yun Li}
}

@article{lCui2023,
    title = {Dynamic pricing for fast charging stations with deep reinforcement learning},
    journal = {Applied Energy},
    volume = {346},
    pages = {121334},
    year = {2023},
    issn = {0306-2619},
    doi = {https://doi.org/10.1016/j.apenergy.2023.121334},
    url = {https://www.sciencedirect.com/science/article/pii/S0306261923006980},
    author = {Li Cui and Qingyuan Wang and Hongquan Qu and Mingshen Wang and Yile Wu and Le Ge}
}

@Article{qChen2023,
AUTHOR = {Chen, Qin and Folly, Komla Agbenyo},
TITLE = {Application of Artificial Intelligence for EV Charging and Discharging Scheduling and Dynamic Pricing: A Review},
JOURNAL = {Energies},
VOLUME = {16},
YEAR = {2023},
NUMBER = {1},
ARTICLE-NUMBER = {146},
URL = {https://www.mdpi.com/1996-1073/16/1/146},
ISSN = {1996-1073},
DOI = {10.3390/en16010146}
}

@ARTICLE{zZhao2022,
  author={Zhao, Zhonghao and Lee, Carman K. M.},
  journal={IEEE Transactions on Transportation Electrification}, 
  title={Dynamic Pricing for EV Charging Stations: A Deep Reinforcement Learning Approach}, 
  year={2022},
  volume={8},
  number={2},
  pages={2456-2468},
  doi={10.1109/TTE.2021.3139674}
  }

@article{bPalaniyappan2024,
title = {Dynamic pricing for load shifting: Reducing electric vehicle charging impacts on the grid through machine learning-based demand response},
journal = {Sustainable Cities and Society},
volume = {103},
pages = {105256},
year = {2024},
issn = {2210-6707},
doi = {https://doi.org/10.1016/j.scs.2024.105256},
url = {https://www.sciencedirect.com/science/article/pii/S2210670724000854},
author = {Balakumar Palaniyappan and Senthil Kumar R and Vinopraba T}
}

@ARTICLE{aNarayan2022,
  author={Narayan, Ajay and Krishna, Aakash and Misra, Prasant and Vasan, Arunchandar and Sarangan, Venkatesh},
  journal={IEEE Intelligent Transportation Systems Magazine}, 
  title={A Dynamic Pricing System for Electric Vehicle Charging Management Using Reinforcement Learning}, 
  year={2022},
  volume={14},
  number={6},
  pages={122-134},
  doi={10.1109/MITS.2022.3198019}
  }

@article{bAljafari2023,
title = {Electric vehicle optimum charging-discharging scheduling with dynamic pricing employing multi agent deep neural network},
journal = {Computers and Electrical Engineering},
volume = {105},
pages = {108555},
year = {2023},
issn = {0045-7906},
doi = {https://doi.org/10.1016/j.compeleceng.2022.108555},
url = {https://www.sciencedirect.com/science/article/pii/S0045790622007704},
author = {Belqasem Aljafari and Pandia Rajan Jeyaraj and Aravind Chellachi Kathiresan and Sudhakar Babu Thanikanti}
}

@ARTICLE{cWei2026,
  author={Wei, Cunhao and Chen, Zhong and Li, Bingrong},
  journal={IEEE Transactions on Transportation Electrification}, 
  title={Mixed Game-Theoretic Scheduling Method for Urban Electric–Hydrogen Integrated Charging Stations Considering Hydrogen Transportation}, 
  year={2026},
  volume={12},
  number={3},
  pages={4497-4512}
}

@ARTICLE{bLi2023,
  author={Li, Bei and Li, Jiangchen and Han, Mei},
  journal={IEEE Access}, 
  title={Deep Reinforcement Learning-Based Charging Price Determination Considering the Coordinated Operation of Hydrogen Fuel Cell Electric Vehicle, Power Network and Transportation Network}, 
  year={2023},
  volume={11},
  number={},
  pages={75508-75521},
  doi={10.1109/ACCESS.2023.3296783}
}

@INPROCEEDINGS{jGbadegoye2025,
  author={Gbadegoye, Jeremiah and Chen, Yang and Omitaomu, Olufemi A. and Li, Xueping},
  booktitle={2025 Winter Simulation Conference (WSC)}, 
  title={Operational Simulation of Multi-Functional Charging Station for Sustainable Transportation}, 
  year={2025},
  volume={},
  number={},
  pages={2308-2319},
  doi={10.1109/WSC68292.2025.11338905}
}

@ARTICLE{sLai2024,
  author={Lai, Shuying and Qiu, Jing and Tao, Yuechuan and Sun, Xianzhuo and Zhao, Junhua},
  journal={IEEE Transactions on Industrial Informatics}, 
  title={Charging/Refueling Navigation Strategies for Plug-in Hybrid Hydrogen and Electric Vehicles With Irrationalities and Energy Substitution}, 
  year={2024},
  volume={20},
  number={1},
  pages={583-595}
}

@ARTICLE{sLai2023,
  author={Lai, Shuying and Qiu, Jing and Tao, Yuechuan and Zhao, Junhua},
  journal={IEEE Transactions on Smart Grid}, 
  title={Pricing for Electric Vehicle Charging Stations Based on the Responsiveness of Demand}, 
  year={2023},
  volume={14},
  number={1},
  pages={530-544},
  doi={10.1109/TSG.2022.3188832}
}

@article{yChen2020,
title = {A comparison study on trading behavior and profit distribution in local energy transaction games},
journal = {Applied Energy},
volume = {280},
pages = {115941},
year = {2020},
issn = {0306-2619},
doi = {https://doi.org/10.1016/j.apenergy.2020.115941},
url = {https://www.sciencedirect.com/science/article/pii/S0306261920313982},
author = {Yang Chen and Byungkwon Park and Xiao Kou and Mengqi Hu and Jin Dong and Fangxing Li and Kadir Amasyali and Mohammed Olama}
}

@article{yChen2025,
title = {Operational optimization for multi-functional charging station with electric and hydrogen-powered vehicles},
journal = {Sustainable Energy, Grids and Networks},
volume = {44},
pages = {101998},
year = {2025},
issn = {2352-4677},
doi = {https://doi.org/10.1016/j.segan.2025.101998},
url = {https://www.sciencedirect.com/science/article/pii/S2352467725003807},
author = {Yang Chen and Jeremiah Gbadegoye and Xudong Wang and Olufemi A. Omitaomu and Xueping Li}
}

@INPROCEEDINGS{xWang2025,
  author={Wang, Xudong and Chen, Yang and Skipper, Brody and Omitaomu, Olufemi A. and Li, Xueping},
  booktitle={2025 Winter Simulation Conference (WSC)}, 
  title={Agent-Based Simulation of Price-Demand Dynamics in Multi-Service Charging Station}, 
  year={2025},
  volume={},
  number={},
  pages={676-687},
  doi={10.1109/WSC68292.2025.11339092}}

@ARTICLE{mAdil2024,
  author={Adil, Muhammad and Mahmud, M. A. Parvez and Kouzani, Abbas Z and Khoo, Sui Yang},
  journal={IEEE Transactions on Industry Applications}, 
  title={Optimal Location and Pricing of Electric Vehicle Charging Stations Using Machine Learning and Stackelberg Game}, 
  year={2024},
  volume={60},
  number={3},
  pages={4708-4722},
  doi={10.1109/TIA.2024.3364579}}

@ARTICLE{hZeng2024,
  author={Zeng, Hongtai and Sheng, Yujie and Sun, Hongbin and Zhou, Yanzhen and Xue, Yixun and Guo, Qinglai},
  journal={IEEE Transactions on Smart Grid}, 
  title={A Conic Relaxation Approach for Solving Stackelberg Pricing Game of Electric Vehicle Charging Station Considering Traffic Equilibrium}, 
  year={2024},
  volume={15},
  number={3},
  pages={3080-3097},
  doi={10.1109/TSG.2023.3329651}}

@article{hWang2024,
title = {Research on energy pricing of the hydrogen refueling station based on master-slave game in multi-market},
journal = {Applied Energy},
volume = {373},
pages = {123906},
year = {2024},
issn = {0306-2619},
doi = {https://doi.org/10.1016/j.apenergy.2024.123906},
url = {https://www.sciencedirect.com/science/article/pii/S0306261924012893},
author = {Haibing Wang and Libo Zhu and Weiqing Sun and Muhammad Qasim Khan and Bin Liu}
}

@article{dMatkovic2025,
title = {Dynamic pricing strategy based on day-ahead charging demand forecasts},
journal = {Sustainable Energy, Grids and Networks},
volume = {44},
pages = {101897},
year = {2025},
issn = {2352-4677},
doi = {https://doi.org/10.1016/j.segan.2025.101897},
url = {https://www.sciencedirect.com/science/article/pii/S2352467725002796},
author = {Daria Matkovic and Terezija {Matijasevic Pilski} and Tomislav Capuder}
}

@Article{dLu2023,
AUTHOR = {Lu, Di and Sun, Jing and Peng, Yonggang and Chen, Xiaofeng},
TITLE = {Optimized Operation Plan for Hydrogen Refueling Station with On-Site Electrolytic Production},
JOURNAL = {Sustainability},
VOLUME = {15},
YEAR = {2023},
NUMBER = {1},
ARTICLE-NUMBER = {347},
URL = {https://www.mdpi.com/2071-1050/15/1/347},
ISSN = {2071-1050}
}

@article{mAnjos2025,
author = {Miguel F. Anjos and Luce Brotcorne and Gaël Guillot},
title = {Optimal electric vehicle charging with dynamic pricing, customer preferences and power peak reduction},
journal = {INFOR: Information Systems and Operational Research},
volume = {63},
number = {3},
pages = {439--458},
year = {2025},
publisher = {Taylor \& Francis},
doi = {10.1080/03155986.2025.2463189},
URL = {https://doi.org/10.1080/03155986.2025.2463189},
eprint = {https://doi.org/10.1080/03155986.2025.2463189}
}

@misc{zevstation2026,
  author       = {{ZEV Station}},
  title        = {Home},
  howpublished = {\url{https://zev-station.com/}},
  year         = {2026},
  note         = {Accessed: Mar. 19, 2026}
}

\end{document}